\documentclass{PoS}
\usepackage{amsmath}
\usepackage[export]{adjustbox}
\usepackage{subcaption}

\title{Constraining the evaporation rate of primordial black holes using archival data from VERITAS}

\ShortTitle{Constraining evaporation rate of PBHs with VERITAS}

\author{\speaker{Sajan Kumar$^{1}$, for the VERITAS collaboration}\thanks{for the VERITAS collaboration.}\\
        $^1$Department of Physics, McGill University, Montreal, Canada\\
        E-mail: \email{sajan.sajankumar@mcgill.ca}}

\abstract{Primordial black holes (PBHs) are thought to have been formed as a result of density fluctuations in the very early Universe. It is suggested that PBHs of mass $\sim 5 \times 10^{14} \mathrm{\ g}$ or less have evaporated through the release of Hawking radiation by the present day. However, PBHs of initial mass $10^{15} \mathrm{\ g}$ should still be evaporating at the present epoch. Over the past few years, very high-energy (VHE; E $>$ 100 GeV) gamma-ray emission from PBHs in the form of a burst has been searched for using ground-based gamma-ray instruments. However, no observational evidence has been reported on the detection of VHE emission from PBHs yet. Previously, an upper limit on the rate density of PBHs was calculated using 750 hours of archival data taken between 2009 and 2012 by the VERITAS gamma-ray observatory. We will augment this study with additional data taken between 2012 and 2017. In addition to more data, the lower energy threshold on the newer data will help to produce an improved upper limit on the rate at which PBHs are evaporating in our local neighborhood. This work is still in progress, therefore we will only report an expected change to the upper limit on the rate density of PBH evaporation.}

\FullConference{36th International Cosmic Ray Conference -ICRC2019-\\
		July 24th - August 1st, 2019\\
		Madison, WI, U.S.A.}

\begin{document}

\section{Introduction}
The observational evidence for the existence of primordial black holes (PBHs) is still lacking. But even their non-detection can lead to important constraints on the physics of the early Universe, high energy particle physics and quantum gravity \cite{Carr2005}. PBHs are theoretical objects that are predicted to be formed due to fluctuations in the density of early Universe \cite{Zeldovich1966, Hawking1971}. These cosmological density fluctuations could have caused the over dense regions to collapse gravitationally, thus leading to the formation of PBHs. The initial mass of a PBH depends upon its creation at the time, $t$, after the Big Bang, according to the equation 1.1 in \cite{Carr2010}. 

\begin{equation}
    M_{BH} \sim 10^{15}\bigg(\frac{t}{10^{-23}\mathrm{\ sec}}\bigg)\mathrm{\ g}
\end{equation}

Thus PBHs can span a large range of masses, from $10^{-5}\mathrm{\ g}$ formed at the Planck time ($10^{-43} \mathrm{\ second}$) to $10^{5} \mathrm{\ M_{\odot}}$ formed at 1 second after the Big Bang.

The presence of such PBHs in the Universe could be confirmed by detecting Hawking radiation (HR) from black holes \cite{HawkingTheory1974}. As the  black hole evaporates, its temperature is given by

\begin{equation}
    kT_{BH} = 1.06 \bigg( \frac {10^{13} \mathrm{\ g}}{M_{BH}}\bigg) \mathrm{\ GeV}
\label{Temp}
\end{equation}

where $k$ is the Boltzmann constant. From the Equation \ref{Temp}, it is clear that the temperature of a black hole is inversely proportional to its mass. As the black hole evaporates, its mass decreases, and therefore the temperature increases. This increase in temperature leads to an increase in emitted particle flux. Therefore, it emits a burst of particles in the last few seconds of its life time. This burst includes gamma-ray photons in the MeV to TeV energy range. The life time of a PBH can be approximated as \cite{Carr2010}  

\begin{equation}
    \tau_{BH} \sim 4.55 \times 10^{-28} \bigg( \frac{M_{BH}}{1 \mathrm{\ g}}\bigg)^{3} \mathrm{\ sec}
\end{equation}

PBHs which were formed in the early universe with an initial mass $\sim 10^{15} \mathrm{\ g}$ would have a lifetime close to the age of the Universe. Therefore, they should be in the final stages of their evaporation at the present time and might be seen using ground based Cherenkov gamma-ray telescopes. In order to verify this, the Very Energetic Radiation Imaging Telescope Array System (VERITAS) has searched for the burst signal from evaporating PBHs with a remaining lifetime window of $1 \mathrm{\ second}$. From the study, a $99\%$ confidence-level upper limit, on the rate-density of PBHs evaporation, has been placed at a value of $1.29 \times 10^{5} \mathrm {\ pc^{-3} yr^{-1}}$ \cite{Tesic2012}. The Milagro experiment also tried to detect a burst signal from evaporating PBHs and calculate an upper limit for $1 \mathrm{\ second}$ time window at a value of $3.6 \times 10^{4} \mathrm {\ pc^{-3} yr^{-1}}$ \cite{Milagro2015}. A similar study was performed by the High Energy Stereoscopic System (H.E.S.S) using a time window of $30 \mathrm{\ seconds}$ to put an upper limit on the PBHs evaporation rate at $1.4 \times 10^{4} \mathrm {\ pc^{-3} yr^{-1}}$ \cite{HESS2013}. More recently, VERITAS obtained their best limits for the rate-density of $2.2 \times 10^{4} \mathrm {\ pc^{-3} yr^{-1}}$ using a time window of $30 \mathrm{\ seconds}$ \cite{VERITAS2017}.  

\section{The VERITAS observatory}
The VERITAS observatory is a system of four ground-based atmospheric Cherenkov telescopes situated in southern Arizona
($31.68 \mathrm{\ N}$, $110.95\mathrm{\ W}$), at an elevation of $1268 \mathrm {\ m}$ above sea level. Each telescope has a $12 \mathrm{\ m}$ diameter optical reflector, providing a total reflecting area of  $\sim 110 \mathrm{\ m^{2}}$. At the focus of each telescope, a pixelated camera consists of 499 photomultiplier tubes (PMTs) is placed giving a field of view of $3.5 \mathrm{\ degree}$ in diameter. Currently, a source with a flux level of 1 \% of steady flux from the Crab Nebula can be detected in 25h. The angular resolution of the array at 1 TeV is $\sim 0.1\mathrm{\ degree}$ and the sensitive energy detection range spans from $100\mathrm{\ GeV}$ to $> 30\mathrm{\ TeV}$ \cite{Nahee2015}.

\section{Data selection and analysis}
This work uses good weather VERITAS archival data from the summer 2009 until the end of 2017. In order to achieve minimal energy threshold for the detection, only data taken with pointing above 50 degree elevation was selected. Furthermore, the selected data after applying the above criterion is divided into two epochs, each depending on the telescope camera configuration. Dataset I consists of data taken after the summer of 2009 and before the summer of 2012. The total useful data between these dates amount to about 750 hours. Dataset II consists of observations taken after the summer of 2012 when the telescope cameras were upgraded with high efficiency photomultiplier tubes resulting in an energy threshold below $100 \mathrm{\ GeV}$. The total useful data after the summer of 2012 until the end of 2017 is approximately 1300 hours. In the total, this study uses about 2100 hours of data to evaluate the limits on the rate-density of PBH evaporation. 

Before applying the PBH search methodology, it is important to remove the background events from the sample of all the events that triggered the telescope readout. This is achieved using the boosted decision tree (BDT) method discussed in \cite{Maria2017}. For the PBH burst search, the previous methodology discussed in \cite{VERITAS2017} is used. However, since the Dataset II is taken with a different telescope configuration, it is necessary to re-evaluate the angular resolution of instrument for Dataset II before applying the the burst search methodology. For the Dataset I, we will be using the same values as already estimated in the previous work \cite{VERITAS2017}.
\subsection{Angular resolution} \label{PSF}
The basic requirement for the definition of a burst is that all the events from a burst should come within a certain time window defined \textit{a priori}. In addition, it is also required that the arrival directions, of all events belonging to a particular burst, should fall within the region defined to match the angular resolution of VERITAS. The calculation for the angular resolution (also called point spread function(PSF)) of the VERITAS instrument is performed by plotting events in the $\theta^{2}$ space, where $\theta$ is a measure of angular distance between the reconstructed event direction and the location of gamma-ray source, and then fitting the  $\theta^{2}$ distribution with a  modified hyperbolic secant distribution following \cite{VERITAS2017}:

\begin{equation}
    S(\theta^{2}, \sigma) = \frac{1.71N}{2\pi \sigma^{2}} sech{(\sqrt {\theta^{2}}/\sigma)}
    \label{fitfunc}
\end{equation}

where $\sigma$ is the width of the distribution, estimated from fitting procedure, representing $55.1\%$ containment radius, and N is the number of signal events.

\begin{figure}[ht]
\begin{center}
\includegraphics [scale=1.0]{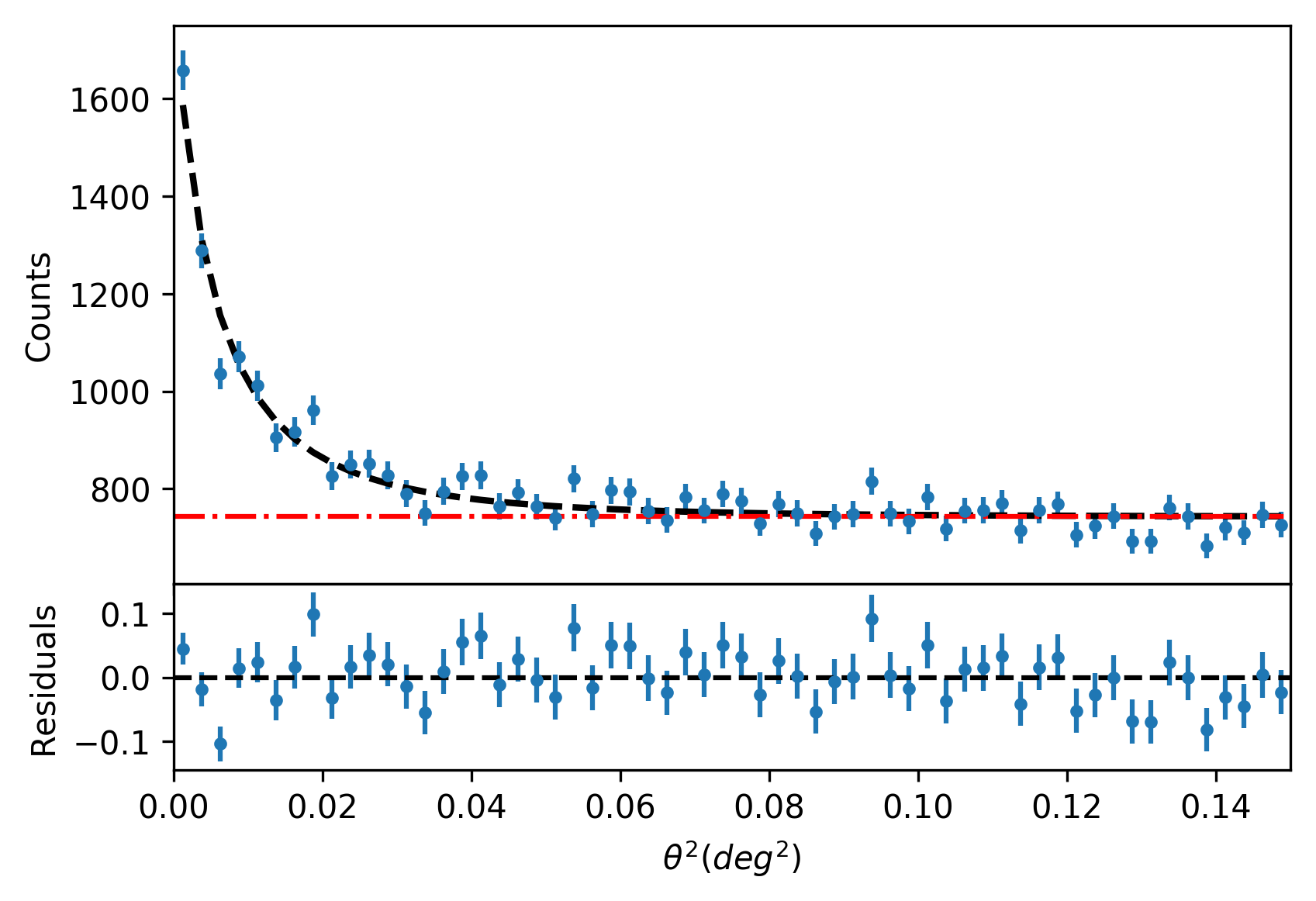}
\end{center}

\caption{\protect \small $\theta^{2}$-distribution of all the events from 3 hours of Crab observations at mean zenith angle of $17 \mathrm{\ degrees}$. The red dotted line represents the constant level of background events and black dotted line represent the signal plus, background profile of the distribution.
}
\label{Theta2}

\vspace{4mm}
\end{figure}

Figure \ref{Theta2} shows the $\theta^{2}$ distribution of events from the Crab Nebulae, fitted with the function defined in Equation \ref{fitfunc} plus a constant function (used to model the background level). Since the PSF of the instrument vary with elevation and energy, separate fits were performed in three elevation bins; 50-$70^{\circ}$, 70-$80^{\circ}$, 80-$90^{\circ}$ and four energy bins; 0.08-0.32 TeV, 0.32-0.5 TeV, 0.5-1 TeV, 1-50 TeV. Table \ref{psftable} shows the values of the $\sigma$ parameter, estimated in different energy and elevation bins using Crab data.

\begin{table*}
\caption{Estimation of dependence of width parameter ($\sigma$) on energy and elevation angle, using Crab data}
 \vspace{5mm}
 \centering
 \begin{tabular}{|c|| c c c|} 
 \hline
  & Elevation & Elevation & Elevation \\ [0.5ex]
   & (50-70 deg) & (70-80 deg) & (80-90 deg) \\ [0.5ex]
 \hline\hline
 Energy & 0.067 $\pm$ 0.006 & 0.068 $\pm$ 0.005 & 0.065 $\pm$ 0.004 \\ 
 (0.08-0.32 TeV) & & &\\
 \hline
 Energy & 0.048 $\pm$ 0.004 & 0.039 $\pm$ 0.003 & 0.040 $\pm$ 0.004 \\
 (0.32-0.5 TeV) & & &\\
 \hline
 Energy & 0.043 $\pm$ 0.003 & 0.045 $\pm$ 0.004 & 0.027 $\pm$ 0.002 \\
 (0.5-1.0 TeV) & & &\\
 \hline
 Eenergy & 0.027 $\pm$ 0.003 & 0.028 $\pm$ 0.003 & 0.028 $\pm$ 0.003 \\
 (1.0-50 TeV) & & &\\
 \hline
\end{tabular}
\label {psftable}
\end{table*}

\subsection{Likelihood method to separate real bursts from background bursts}
\begin{figure}[h]
 
\begin{subfigure}{0.5\textwidth}
\includegraphics[width=0.7\linewidth, height=5cm]{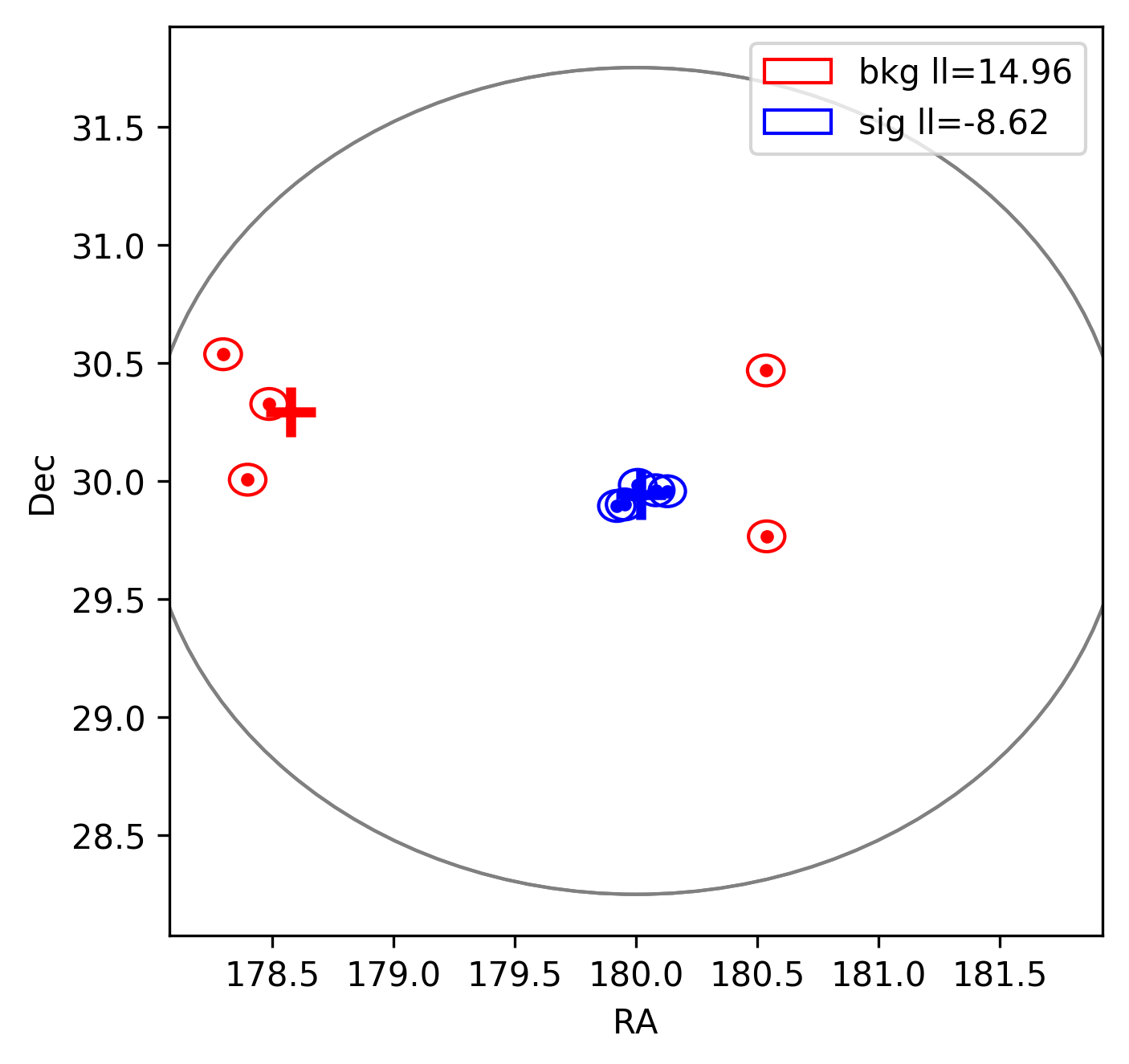} 
\label{fig:subim1}
\end{subfigure}
\begin{subfigure}{0.5\textwidth}
\includegraphics[width=0.9\linewidth, height=5cm]{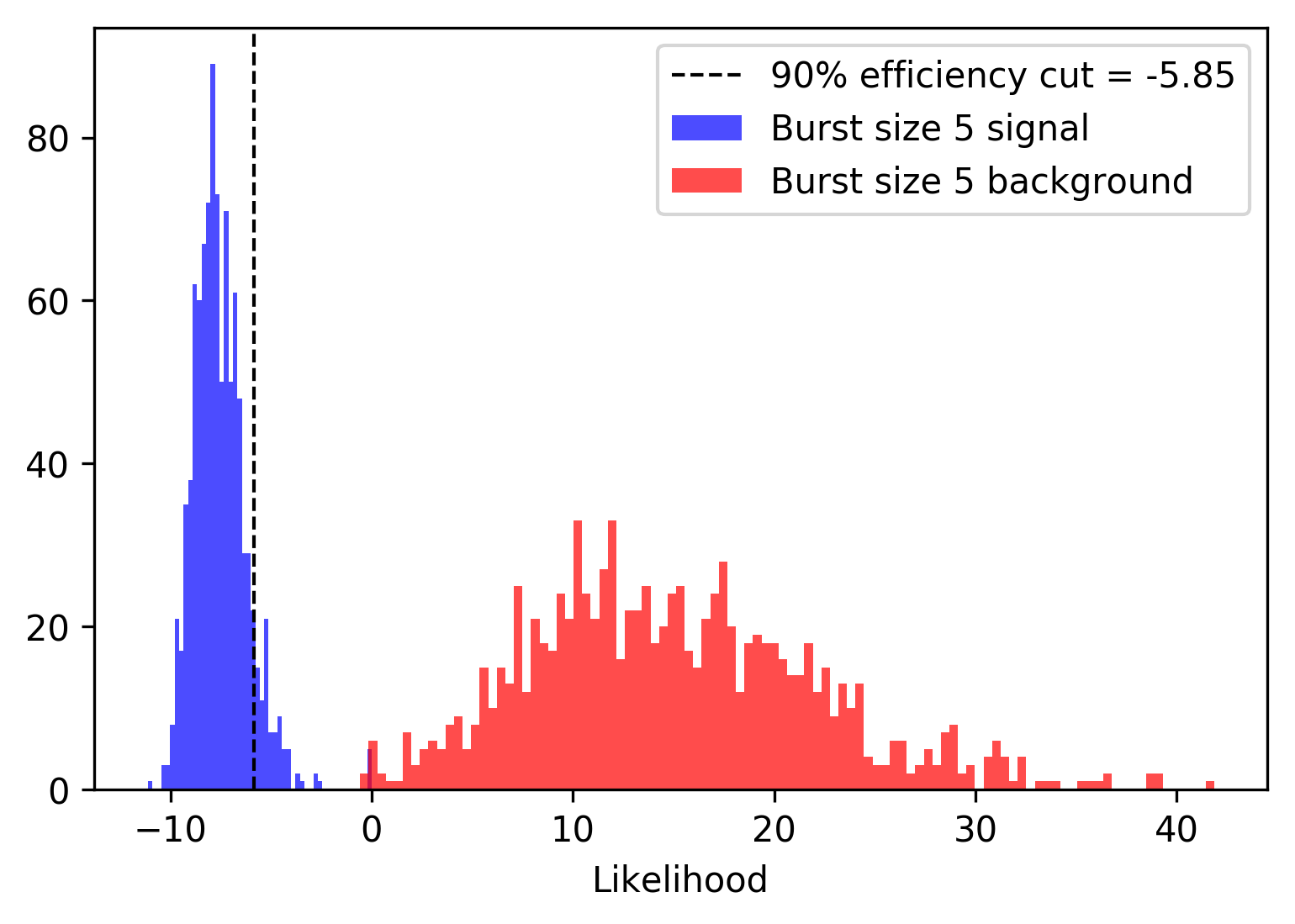}
\label{fig:subim2}
\end{subfigure}
 
\caption{\textit{Left}: Event centroid for a random group of events (red) and a group of events whose positions were randomly generated from Equation \ref{fitfunc}, to simulate a real burst (blue). The circle represents the field of view of VERITAS. \textit{Right}: Distribution of simulated signal bursts compared with random groups, of 5 events each, to determine cut on likelihood value to use in the analysis. The black-dotted vertical line represents the cut value for likelihood that includes $90\%$ of the signal. }
\label{fig:image2}
\end{figure}

In order to separate a real burst of gamma-rays, coming from a point source in the sky, from a set of background coincidental events that appear to mimic a burst, a likelihood method is employed as explained in \cite{VERITAS2017}. The likelihood function  for the set of events whose arrival directions are contained within the angular resolution of the camera ($\sim 0.1 \mathrm{\ degree}$) can be written as:
\begin{equation}
L = \prod_{i} \frac{1.71N}{2\pi \sigma_{i}^{2}} sech{(\sqrt {(\theta_{i} - \mu)^{2}}/\sigma_{i})}
\label{ll}
\end{equation}

where $\sigma_{i}$ and $\theta_{i}$ are the width parameter and direction of event $i$ respectively. The width can be derived from the Table \ref{psftable} for a given event falling in a particular energy and elevation bin. $\mu$ represents the centroid for the given set of events. Figure \ref{fig:image2} (\textit{Left}) shows the centriod found using Equation \ref{ll} from a set of randomized events that represents background (in red), and centroid of group of events generated randomly according to the Equation \ref{fitfunc} that represents a simulated signal (in blue). 

Figure \ref{fig:image2} (\textit{Right}) shows a likelihood distribution for groups of 5 events coming from background bursts (red curve) and simulated signal bursts (blue curve). In order to maximize the amount of signal bursts and minimize the contamination from background bursts, a cut on the likelihood value is determined which retains $90\%$ of the signal bursts. The same procedure is repeated for a burst size of 2, 3, 4, up to 10 events, after which the likelihood of finding bursts with so many events becomes low, and the cuts tend to stay fairly constant between burst sizes. 
 
\section{Burst search method}
For each run, consisting of $\sim$ 30 minutes of observations, a list of gamma-like events (events that passes gamma-hadron separation cuts) is created. For each gamma-like event arriving at time $t_{i}$, a list is then compiled with any subsequent events arriving within a time window of $1, 2, 3, 5, 10, 20, 30 \mathrm{, \ and \  45} \mathrm{\ seconds}$.  

\begin{figure}[ht]
\begin{center}
\includegraphics [scale=0.4]{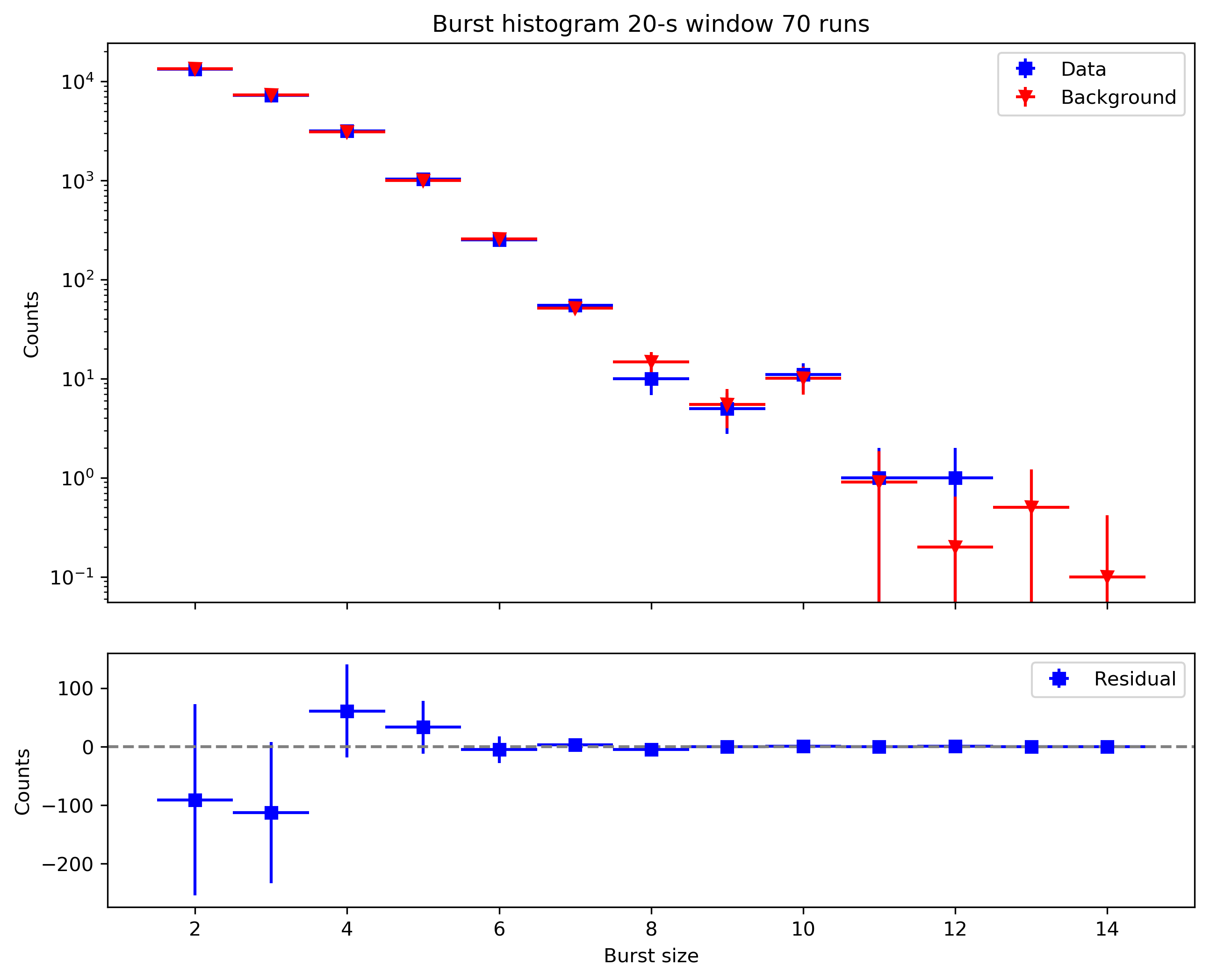}
\end{center}

\caption{\protect \small Burst distribution for a burst duration of $20 \mathrm{\ seconds}$. Top: The blue squares are from the
data, and the red triangles are the background estimation. Bottom: Residual plot showing the difference between data and
background.
}
\label{bursthist}

\vspace{4mm}
\end{figure}

In order to estimate the expected background rate due to chance coincidence, the method in \cite{Linton2006} is used. This methodology consists of taking all the gamma-like events in a data run, and scrambling their time of arrival while keeping their arrival direction unchanged. Any bursts found in this scrambled data can be treated as background. This scrambling procedure is then repeated ten times. By taking an average of these ten iterations, the average background rate can be estimated. For example, for a time-window of $20 \mathrm{\ seconds}$, Figure \ref{bursthist} shows a burst distribution found using this burst search methodology described in detail in \cite{VERITAS2017}. The top plot compares the data points with the background estimation, while the lower plot shows the residuals.

\section{Model Predictions and expectations}
The number of expected photons,  detected by the VERITAS depends upon the emission model of evaporation of PBH and the effective area of the instrument. Following the methodology of \cite{Milagro2015}, we consider a time-integrated spectrum from an evaporation of a PBH for a given value of $\tau$ (remaining life time of PBH) as:
\begin{equation}
    \frac{dN_{\gamma}}{dE_{\gamma}} \approx 9 \times 10^{35} \times \left\{\begin{array}{@{}ll@{}}
    \big(\frac{1 GeV}{T_{\tau}}\big)^{3/2} \big(\frac{1 GeV}{E_{\gamma}}\big)^{3/2} \ \mathrm{GeV^{-1}} & \text{for}\ E_{\gamma} < kT_{\tau} \\
    \big(\frac{1 GeV}{E_{\gamma}}\big)^{3} \ \mathrm{GeV^{-1}} & \text{for}\ E_{\gamma} \geq kT_{\tau}
  \end{array}\right.
\label{spectrum}
\end{equation}

where $T_{\tau}$ defined as $kT_{\tau} = 7.8(\tau/1s)^{-1/3} \mathrm{\ TeV}$, is the temperature of the black hole at the beginning of the final burst time interval. By convolving this model spectrum with the detector response, expected number of gamma-ray events, $N_{\gamma}$, can be estimated as:
\begin{equation}
    N_{\gamma} (r,\alpha, \delta, \tau) = \frac{1}{4\pi r^{2}} \int_{E_{thresh}}^{\infty} \frac{dN_{\gamma}}{dE_{\gamma}}(E_{\gamma},\tau) \ A(E_{\gamma}, \theta_{z},\theta_{w},\mu,\alpha,\delta)\ dE_{\gamma}
\end{equation}

where $E_{thresh}$ is the threshold energy of VERITAS array, and $A(E_{\gamma}, \theta_{z},\theta_{w},\mu,\alpha,\delta)$ is the detector response function (effective area and camera
radial acceptance) of VERITAS, as a function of the gamma-ray energy $E_{\gamma}$, the observation zenith angle $\theta_{z}$, the offset of source from the center of the camera $\theta_{w}$, the optical efficiency $\mu$, and the event reconstruction position in camera coordinates $(\alpha,\delta)$.

The probability of seeing a burst of a certain number of photons $b$ from a PBH emitting $N_{\gamma}$  VHE
photons is expressed through Poisson statistics: $P(b,N_{\gamma}) = e^{-N_{\gamma}}N_{\gamma}^{b} / b!$. This then is tranlated into effective detectable volume as:
\begin{equation}
    V_{eff}(b,\tau) = \int_{0}^{\Delta \Omega} \int_{0}^{\infty} drr^{2}P(b, N_{\gamma})
\end{equation}

The expected number of bursts of size $b$ seen by VERITAS over a total time period of observations $T_{obs}$ can be written as:
\begin{equation}
    n_{exp} (b,\tau) = \dot{\rho}_{PBH} \times T_{obs} \times V_{eff}(b,\tau)
    \label{ratedensity}
\end{equation}

where $\dot{\rho}_{PBH}$  is the rate-density of PBH evaporation. From Equation \ref{ratedensity}, an upper limit on the rate density can be calculated by estimating the limit on the number of expected events. 

The previous reported limit, calculated using 747 hours of data belongs to Dataset I is given by $2.22 \times 10^{4} \mathrm{\ pc^{-3}yr^{-1}}$. The new limit, which is still a work in progress, is expected to be a factor of two more constraining from the previous limit, at a value of about $10^{4} \mathrm{\ pc^{-3}yr^{-1}}$.

\section*{Acknowledgements}
This research is supported by grants from the U.S. Department of Energy Office of Science, the U.S. National Science Foundation and the Smithsonian Institution, and by NSERC in Canada. This research used resources provided by the Open Science Grid, which is supported by the National Science Foundation and the U.S. Department of Energy’s Office of Science, and resources of the National Energy Research Scientific Computing Center (NERSC), a U.S. Department of Energy Office of Science User Facility operated under Contract No. DE-AC02-05CH11231. We acknowledge the excellent work of the technical support staff at the Fred Lawrence Whipple Observatory and at the collaborating institutions in the construction and operation of the instrument.

\bibliographystyle{unsrt}
\bibliography{listofreferences.bib}

\end{document}